# *Saturable absorption versus 'slow light'*

**A C SELDEN**

Department of Physics University of Zimbabwe

MP 167 Mount Pleasant HARARE Zimbabwe

e-mail address *acselden@science.uz.ac.zw*

ABSTRACT

Quantitative evaluation of some recent 'slow light' experiments based on coherent population oscillations (CPO) shows that they can be more simply interpreted as saturable absorption phenomena. Therefore they do not provide an unambiguous demonstration of 'slow light'. Indeed a limiting condition on the spectral bandwidth is not generally satisfied, such that the requirements for burning a narrow spectral hole in the homogeneously broadened absorption line are not met. Some definitive tests of 'slow light' phenomena are suggested, derived from analysis of phase shift and pulse delay for a saturable absorber

Introduction

It has recently been pointed out that saturable absorption can explain many of the features attributed to 'slow light' in a host medium, such as delayed pulse transmission and phase shift of an optical signal, without assuming hole burning via coherent population oscillations (CPO) or an associated reduction in group velocity [1]. Similar considerations apply to pulse propagation in GaAs quantum wells [2] and InAs quantum dots [3]. We give a brief outline of signal transmission by a saturable absorber and make quantitative comparison with the results of some recent 'slow light' experiments [4, 5, 6] with a view to resolving this issue

Saturable absorption

For an incident pulse of width $\tau_p \sim \tau_s$, the lifetime of the metastable state in a saturable absorber, the time dependent transmission lags behind the intensity variation of the incident pulse, resulting in pulse distortion and delayed peak transmission [7-10], while a long pulse (step function) of sufficient intensity generates a 'bleaching wave' with a propagation velocity of a few metres per second [11]. For a weakly modulated input, signal distortion is negligible, the transmitted intensity acquiring an enhanced modulation depth and frequency dependent phase shift, as observed in ruby [4], $Er^{3+}:Y_2SiO_5$ [5] and bacteriorhodopsin [6]. However, the phase shift merely reflects the finite response time of the saturable absorber [12, 13] rather than a change in group velocity [4, 5, 6]

For an optically thick saturable absorber with a weakly modulated (m<<1) incident intensity $I_{in}(t) = I_0[1+m\exp(i\omega t)]$, small signal analysis shows that the transmitted intensity $I_{out}(t) = I_0 T_s[1+mK(\omega)\exp(i\omega t)]$, where $T_s$ is the steady state transmission, given by $\ln(T_s/T_0) = \beta(1-T_s)$, $T_0$ is the initial transmission, parameter $\beta = I_0/I_s$, where $I_s$ is the saturation intensity and the modulation factor [14, 15]

$$K(\omega) = \frac{1+\beta+i\omega\tau_s}{1+\beta T_s + i\omega\tau_s} \qquad (1)$$

Thus the transmitted signal has modulation gain $|K(\omega)| > 1$ and phase $\varphi = \arg K(\omega) < 0$. Comparing $|K(\omega)|$ from eq (1) with the relative modulation attenuation $A(\omega)$ defined in [4], we find $A(\omega) \equiv -\ln|K(\omega)|$, with a maximum value $A(0) = \ln\{(1+\beta T_s)/(1+\beta)\}$ at zero modulation frequency $\omega = 0$. Thus saturable absorption theory gives the same result as the theory of coherent population oscillations (CPO) for non-linear signal transmission [15]. Maximum phase lag $\varphi_m$ occurs when $\omega = \omega_m$, where $\omega_m^2\tau_s^2 = (1+\beta)(1+\beta T_s)$ [14], while the maximum delay in signal transmission $\tau_d = \varphi/\omega$ is reached at zero modulation frequency: $\tau_d|_{\omega=0} = \tau_s\beta(1-T_s)/(1+\beta)(1+\beta T_s)$, with a limiting value $\tau_d \Rightarrow \tau_s\beta/(1+\beta) \Rightarrow \tau_s$ for $T_s<<1$ and $\beta \Rightarrow \infty$ i.e. the signal delay never exceeds the lifetime of the metastable state [15]. A similar analysis for gain saturation in an optical amplifier predicts a forward phase shift and temporal advance in signal transmission: $\tau_a|_{\omega=0} = \tau_s\beta(G_s-1)/(1+\beta)(1+\beta G_s)$, where $G_s$ is the steady state gain at intensity $I_0$ and $\ln(G_0/G_s) = \beta(G_s-1)$. Thus $\tau_a \Rightarrow \tau_s$ as $G_s \Rightarrow \infty$

A C Selden    22$^{nd}$ March 2006

Comparison with experiment

Fig 1 shows the frequency dependence of the relative modulation attenuation $A(\omega)$ and delay $\tau_d(\omega)$ for a weakly modulated signal transmitted by a saturable absorber. The theoretical curves were derived using the parameters for ruby and show very similar characteristics to the experimental data for 'slow light' [4]. Fitting to the datum $\tau_d = 612$ μs at 60 Hz [4] gives near perfect agreement, as does the plot of $A(\omega)$ vs. modulation frequency, interpreted as a 'spectral hole' in the absorption profile [4, 12]. Similar results are found for $^{167}Er^{3+}$:$Y_2SiO_5$ [5], as shown in Fig 2, where the variation of signal delay with incident intensity for modulation frequencies 0, 10 and 40 Hz is plotted, assuming saturable absorption. The curves are of similar form, the higher frequency having a flatter response and lower maximum, as observed. The experimental points for 10 Hz follow the theoretical curves up to a peak value at $I_0 \approx 0.7 I_s$ and down to $I_0 \approx I_s$, diverging thereafter to a long tailed distribution due to inhomogeneous broadening [5, 16]. Saturable absorption also predicts a modulation gain $K = 1.55$ at 10 Hz, in precise agreement with the observed 55% increase in sideband transmission [5]. 'Slow light' implies signal delay $\tau_d \propto \alpha_0 L$ [4, 5] and both saturable absorption and hole burning theory predict virtually identical behaviour for $Er^{3+}$:$Y_2SiO_5$. Calculation shows a near linear dependence of $\tau_d$ on sample length L, assuming saturable absorption, and a minimum apparent 'signal velocity' $V_s = L/\tau_d = 2.74$ m/s for $L = 3$ mm, in agreement with the experimental value $2.7 \pm 0.2$ m/s derived from the maximum observed delay $\tau_d = 1.1 \pm 0.1$ ms at 10 Hz [5].

Turning now to active media with controllable characteristics, such as fibre optic amplifiers and semiconductor waveguides, whose gain (or loss) can be varied by adjusting the optical pumping intensity or bias current, we may apply non-linear transmission theory to the whole range of operation, from saturable absorption to gain saturation [3, 6, 7, 15, 17]. For example, the zero frequency time delay/advance for absorption/gain saturation in a non-linear medium is plotted in Fig 3, showing good agreement with the rescaled experimental data for pulse transmission in a quantum dot amplifier [3]. The time delay peaks slightly above saturation intensity ($\beta = 1.25$) at $\tau_d = 0.4\tau_s$ for $T_0 = 0.05$ ($G_0 = -13$ dB), while the time advance peaks a little below saturation ($\beta = 0.3$) at $\tau_a = 0.2\tau_s$ for $G_0 \approx 6$ dB. In the case of bacteriorhodopsin (bR), a photosynthetic biomolecule characterised by saturable absorption of yellow-green light (570 nm), the relaxation rate can be controlled by illumination with blue light [6]. Increasing the intensity $I_b$ of blue light ($\lambda = 442$ nm) reduces the excited state lifetime as $\tau_s = \tau_1/(1+AI_b)$, where $A = 1.18$ mW$^{-1}$ and $\tau_1 = 0.28$ sec [6]. The dependence of the signal delay on $I_b$ for these parameters is shown in Fig 4, saturable absorption theory showing excellent agreement with the experimental data.

Discussion

One of the problems with 'slow light' experiments is that they do not offer any direct evidence of reduced group velocity, this being inferred from observations of phase shift and pulse delay [4, 5, 6], which can be equally interpreted as saturable absorption phenomena [1-3, 10-15]. For pulse transmission, a more convincing demonstration of 'slow light' requires the delay to exceed the pulse width, ideally for distortion free transmission [18]; the maximum delay reported is approximately 40% of the pulse width, but with considerable distortion through absorption of the leading edge [3]. In fact, saturable absorption significantly reduces the pulse width [7-10], whereas group velocity dispersion in a 'slow light' medium can only increase it [19]. For amplitude modulated transmission, demonstration of a controllable phase shift $\varphi \geq \pi/2$ or delay $\tau_d > \tau_s$ should suffice, because these exceed the limits set by saturable absorption [15]. Here we have seen that 'slow light' effects in ruby, $Er^{3+}:Y_2SiO_5$ and bacteriorhodopsin can be quantitatively evaluated in terms of saturable absorption rather than hole burning and group velocity reduction via CPO. A similar mechanism (gain saturation) has been suggested for slow (and fast) light effects observed in a quantum dot amplifier [3]. Separation of the pump and probe beams would seem a minimum requirement for distinguishing between hole burning and saturable absorption e.g. via transverse excitation and time delayed axial probe [19]. However, the spectral width $\Delta\omega_p$ of laser sources employed in 'slow light' experiments performed with amplitude modulated light has generally exceeded the width of the coherent hole $\Delta\omega_h \sim 1/\tau_s$ by some orders of magnitude; typically $\Delta\omega_p \tau_s \geq 100$-$1000$ [4, 5, 6], such that the minimum condition for burning a narrow hole in the homogeneous line – essential for observing slow light – has not been met [1]. In the linear dispersion regime the delay $\tau_d$ varies inversely with spectral width [6]. Thus any 'slow light' effect would be negligible compared with saturable absorption, the dispersion $\partial n/\partial \omega$ being $\sim 100$-$1000$ times less than assumed. This is particularly true for relaxation times $\tau_s \geq 1s$, as in bR [6], further favouring a saturable absorption model [13]. It is only when a sufficiently narrow line source, such as a tunable diode laser, is used to scan the absorption spectrum of a sample resonantly excited within the homogeneous linewidth that a hole can in fact be observed [20]. These considerations do not conflict with the interpretation of amplitude modulation spectroscopy performed with a broad-band source, which is adequately described by saturable absorption theory [12, 13], but do bring into question the observation of 'slow light' with this technique. The fallacy lies in interpreting the delayed response of a saturable absorber as a signal transit time and using this to derive an apparent 'group velocity' $V_g = L/\tau_d \geq L/\tau_s$, which can be made arbitrarily small (and the 'group index' arbitrarily large) for an indefinitely thin sample ($L \Rightarrow 0$) of given optical density $\alpha_0 L$. Indeed this error is repeated in 'slow light' experiments utilising other physical effects [1]. A strong indication of the true situation is that in most cases the observed delay never exceeds the relaxation time, and more often is just a fraction of it [4, 5, 6], whereas the transit time for slow light should clearly depend on sample length, and readily exceed the response time.

Conclusion

The results of 'slow light' experiments based on transmission of amplitude modulated light agree precisely with saturable absorption theory. As such they do not provide sufficient demonstration of a reduced group velocity in the medium being probed. Further, a limiting condition on the spectral width of the pump source is often violated by some orders of magnitude, such that the experimental requirements for burning a coherent hole $\Delta\omega_h \sim 1/\tau_s$ within the homogeneous linewidth are not met. It is only when this is taken into account that more credible 'slow light' experiments utilising coherent population oscillations (CPO) can be designed and carried out [20]

Acknowledgement

My thanks to Mike van der Poel of the Nanophotonics Group at DTU Lyngby for providing a number of recent publications on pulse propagation in quantum devices

Figure captions

Fig 1   Time delay $\tau_d$, phase shift $\varphi(\omega)$ and relative modulation attenuation $A(\omega)$ of the transmitted intensity vs. modulation frequency, comparing saturable absorption theory with the experimental data for 'slow light' in ruby [4]

Fig 2   Comparison of saturable absorption theory with experimental data for time delay $\tau_d$ vs. intensity ratio $\beta = I_0/I_s$ in $Er^{3+}:Y_2SiO_5$ at 0, 10 and 40 Hz modulation frequency [5]. The dashed curve shows the effect of inhomogeneous broadening [16]

Fig 3   Maximum time delay/advance vs. $\beta$ for absorption/gain saturation in a quantum dot semi-conductor amplifier. The curves are labelled with the unsaturated loss/gain in dB. The points are taken from the experimental data for short pulse transmission [3] with rescaled saturation parameter

Fig 4   Bacteriorhodopsin thin film signal delay $\tau_d$ vs intensity $I_b$ of blue light, which controls the metastable M state lifetime. The points are the experimental data of Wu and Rao [6], the curve follows from eq (1), with intensity dependent lifetime $\tau_s(I_b)$

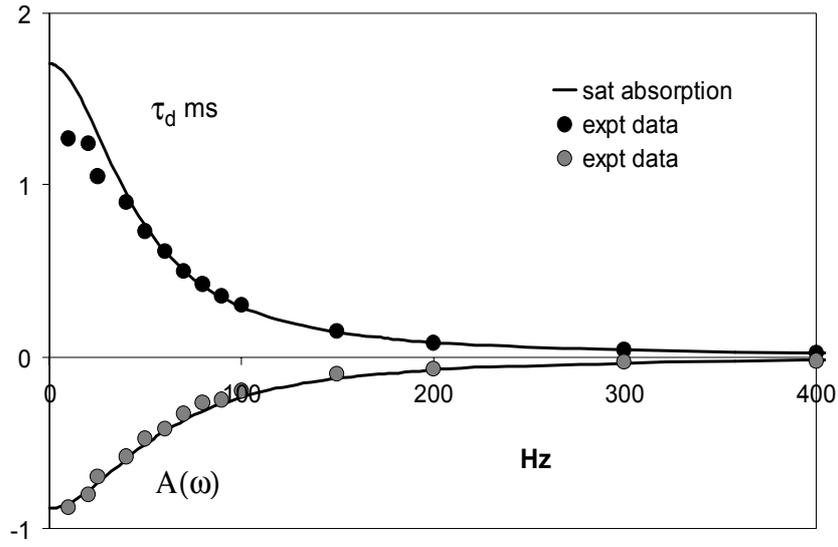

Fig 1

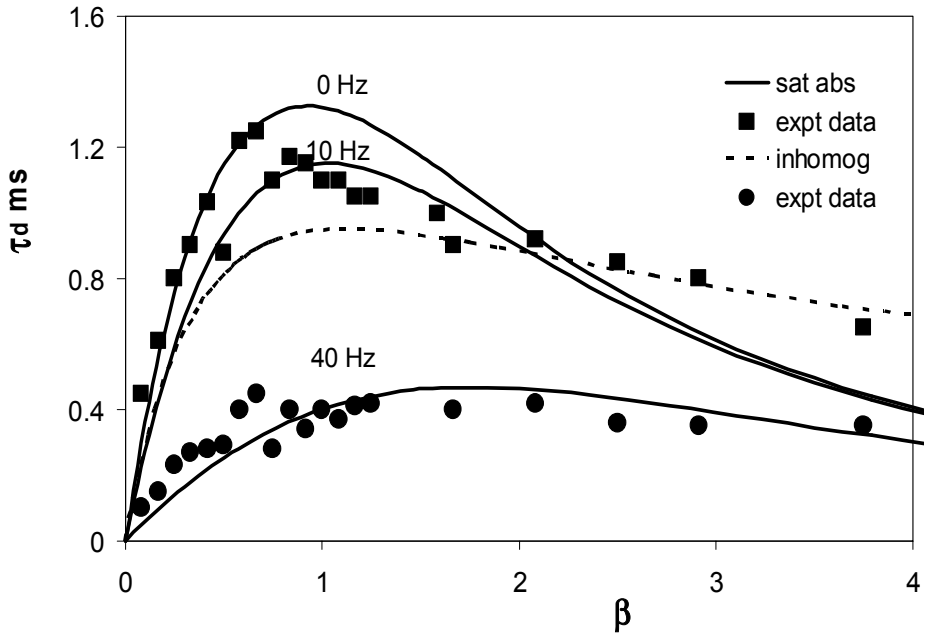

Fig 2

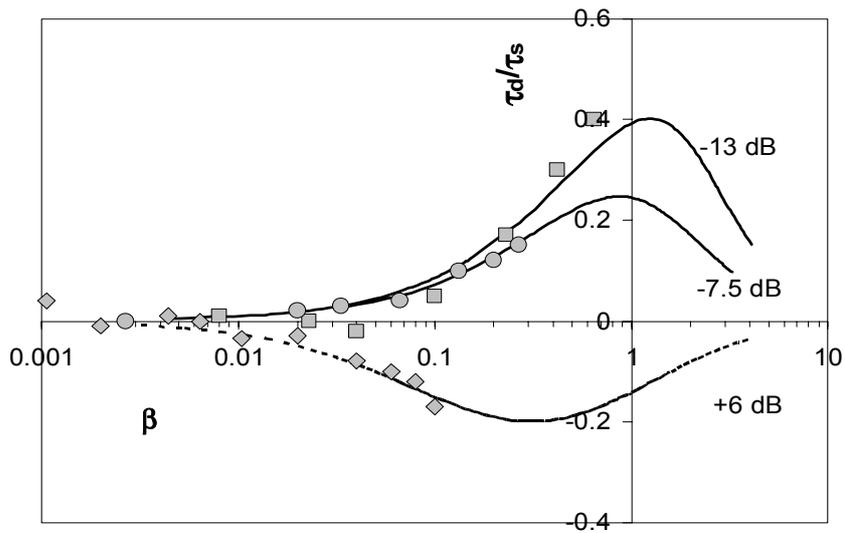

Fig 3

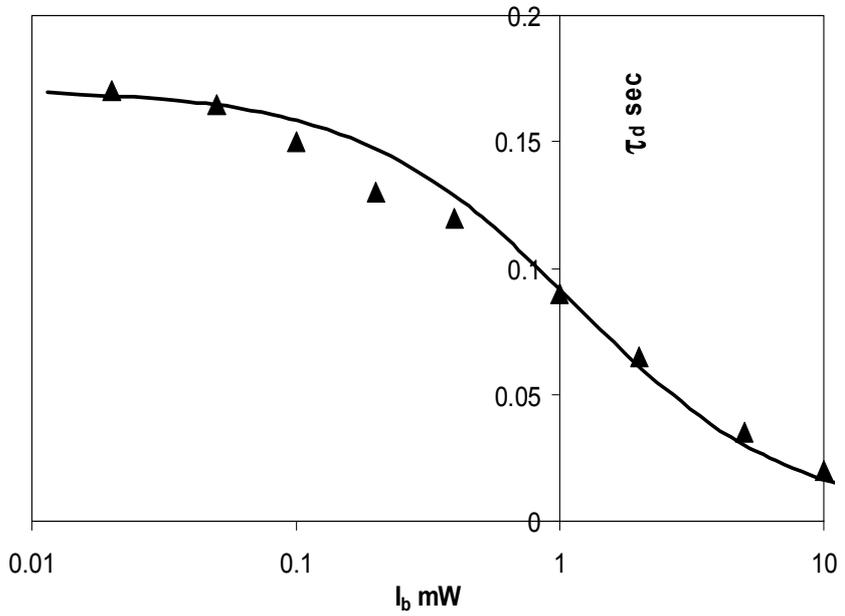

Fig 4